\documentclass[12pt]{article}

\usepackage{epsf,amsfonts,hyperref}
\renewcommand{\appendix}[1]{
    \addtocounter{section}{1}
    \setcounter{equation}{0}
    \renewcommand{\thesection}{\Alph{section}}
    \section*{Appendix \thesection\protect\indent #1}
    \addcontentsline{toc}{section}{Appendix \thesection\ \ \ #1}
}
\newcommand\encadremath[1]{\vbox{\hrule\hbox{\vrule\kern8pt
\vbox{\kern8pt \hbox{$\displaystyle #1$}\kern8pt}
\kern8pt\vrule}\hrule}}
\def\enca#1{\vbox{\hrule\hbox{
\vrule\kern8pt\vbox{\kern8pt \hbox{$\displaystyle #1$}
\kern8pt} \kern8pt\vrule}\hrule}}

\newcommand\figureframex[3]{
\begin{figure}[bth]
\hrule\hbox{\vrule\kern8pt
\vbox{\kern8pt \vbox{
\begin{center}
{\mbox{\epsfxsize=#1.truecm\epsfbox{#2}}}
\end{center}
\caption{#3}
}\kern8pt}
\kern8pt\vrule}\hrule
\end{figure}
}
\newcommand\figureframey[3]{
\begin{figure}[bth]
\hrule\hbox{\vrule\kern8pt
\vbox{\kern8pt \vbox{
\begin{center}
{\mbox{\epsfysize=#1.truecm\epsfbox{#2}}}
\end{center}
\caption{#3}
}\kern8pt}
\kern8pt\vrule}\hrule
\end{figure}
}

\newcommand{\beq}{\begin{equation}}
\newcommand{\eeq}{\end{equation}}
\newcommand{\bea}{\begin{eqnarray}}
\newcommand{\eea}{\end{eqnarray}}

\renewcommand{\and}{{\qquad {\rm and} \qquad}}

\newcommand{\virg}{{\qquad , \qquad}}

 \newcommand{\Tr}{{\,\rm Tr}\:}

\newcommand{\ee}[1]{{{\rm e}^{#1}}}

\renewcommand{\d}{{{\partial}}}

\newcommand{\Pint}{{\int\kern -1.em -\kern-.25em}}

\renewcommand{\Re}{{\mathrm{Re}}}
\renewcommand{\Im}{{\mathrm{Im}}}

\newcommand{\acycle}{{\cal A}}
\newcommand{\bcycle}{{\cal B}}
\newcommand{\genus}{{\overline{g}}}

\renewcommand{\L}{\Lambda}

\newcommand{\bfeps}{{\mathbf \epsilon}}

\newcommand{\bfn}{{\mathbf n}}
\newcommand{\bfzeta}{{\mathbf \zeta}}

\begin{document}
\sloppy


\pagestyle{empty}
\hfill SPhT-T08/028
\addtolength{\baselineskip}{0.20\baselineskip}
\begin{center}
\vspace{26pt}
{\large \bf {Large N expansion of convergent matrix integrals, holomorphic anomalies, and background independence}}
\newline
\vspace{26pt}

{\sl B.\ Eynard}\hspace*{0.05cm}\footnote{ E-mail: eynard@spht.saclay.cea.fr },
\vspace{6pt}
Institut de Physique Th\'{e}orique de Saclay,\\
F-91191 Gif-sur-Yvette Cedex, France.\\
\end{center}

\vspace{20pt}
\begin{center}
{\bf Abstract}:
\end{center}

We propose an asymptotic expansion formula for matrix integrals, including oscillatory terms (derivatives of theta-functions) to all orders.
This formula is heuristically derived from the analogy between matrix integrals, and formal matrix models (combinatorics of discrete surfaces), after summing over filling fractions.
The whole oscillatory series can also be resummed into a single theta function.
We also remark that the coefficients of the theta derivatives, are the same as those which appear in holomorphic anomaly equations in string theory, i.e. they are related to degeneracies of Riemann surfaces. Moreover, the expansion presented here, happens to be independent of the choice of a background filling fraction.

%





\vspace{26pt}
\pagestyle{plain}
\setcounter{page}{1}



\vfill\eject
\section{Introduction}

Convergent matrix integrals of the form
\beq
\hat{Z}=\int_{H_n} dM\,\, \ee{-N\Tr V(M)}
\eeq
are very usefull in many areas of physics (statistical physics, mesoscopic physics, quantum chaos,...) and in mathematics (probabilities, orthogonal polynomials,...) \cite{Mehta, courBleher}.
People are mostly interested in their asymptotic behavior in the large $n$ limit (and $n/N \sim$ constant).
\smallskip

There is another form of matrix integrals, called formal-matrix integrals, which come from combinatorics (2d quantum gravity for physicists \cite{BIPZ, ZJDFG, eynform}).
They are generating functions for counting discrete surfaces (also called "maps") of given topology.
Formal matrix integrals are only asymptotic series, they are not convergent in general, and almost by definition, they always have a large $n$ expansion (see \cite{eynform}).
All the terms in their large $n$ expansion are known \cite{eynloop1mat, EOFg}, and are deeply related to algebraic geometry and integrable systems.
They have many applications to combinatorics, and string theory in physics \cite{mmhouches, vonk}.

\medskip
In this article, we use the analogy between the two types of matrix integrals, and generalizing the method of \cite{BDE}, we propose an asymptotic formula for convergent matrix integrals, including oscillations to all orders:
\bea\label{mainformulaintro}
\hat{Z} 
&\sim& \ee{\sum_g N^{2-2g}F_g}\, \,\left( \Theta + {1\over N}(F_1'\Theta'+{F_0'''\over 6}\Theta''') + \dots \right) \cr
&\sim&  \ee{\sum_g N^{2-2g}F_g}\, \,\,  \sum_{k} \sum_{l_i}\sum'_{h_i} {N^{\sum_i (2-2h_i-l_i)}\over k! l_1!\,\dots\, l_k!} F_{h_1}^{(l_1)}\dots F_{h_k}^{(l_k)}  \,\,\partial^{\sum l_i} \Theta
\cr
\eea
where $\Theta$ is a theta function, i.e. a periodic function, this is why we call $\Theta$ and its derivatives "oscillatory terms".

\medskip

Then we observe that the series containing the oscillatory terms can be resummed into a single theta function:
\beq
\hat{Z}
\sim  \ee{\sum_g N^{2-2g}F_g}\, \,\,\, \Theta\left(NF_0'+\sum_{k=1}^\infty N^{1-2k} u^{(k)}, i\pi \tau + \sum_{j=1}^\infty N^{-2j} t^{(j)}\right) 
\eeq

\medskip
We also observe that the coefficients in front of the derivatives of $\Theta$ in eq.\ref{mainformulaintro}, are the same which appear in the so-called "holomorphic anomaly equations" discovered in the context of topological string theory \cite{BCOV}.
In other words they are related to the combinatorics of degeneracies of Riemann surfaces.

\medskip
Finally, we observe, that although we define each term of the expansion after choosing a reference filling fraction $\bfeps^*$, the partition function is in fact independent of that choice. This is related to the so-called background independence problem in string theory, first observed by Witten \cite{Witten}.

\bigskip

For the 1-hermitian matrix model (with real potential), the first term of this asymptotic expansion 
\beq
\hat{Z}  \sim \ee{\sum_g N^{2-2g}F_g}\, \, \Theta 
\eeq
was derived rigorously by Deift\& co \cite{DKMVZ} using Riemann-Hilbert methods, and their method proved the existence of a whole oscillatory series containing derivatives of the Theta-function.
The same result was also obtained by heuristic physicists methods by \cite{BDE}.
Here, we generalize the method of \cite{BDE} and we give the exact coefficient of the whole series.

Also, in case where the genus of the Theta function is zero, there is no oscillatory term, and one finds the so-called topological expansion $\hat{Z}  \sim \ee{\sum_g N^{2-2g}F_g}$, which is well known to coincide (in the sense of asymptotic formal series) with the generating function for enumerating discrete surfaces \cite{BIPZ}.
In this genus zero case, the asymptotics of the convergent matrix integral were derived by several methods and several authors \cite{EML, Guionnet}.
The coefficient of the expansion are of course the symplectic invariants of \cite{EOFg}.

\medskip

For other convergent matrix models, for instance the 2-matrix model, such expansions were conjectured many times \cite{eynhabilit, eynchain}, but never proved.
Here, we don't prove it either. We merely give all the coefficients in the formula to prove, and we explain their heuristic origin.

As we said above, the heuristic origin of the formulae presented in this article, is just the analogy between formal and convergent matrix models.

\bigskip
{\bf Outline:}

$\bullet$ In the first section, we define the convergent matrix model on generalized paths, and write it as a sum over filling fractions.

$\bullet$ In the second section, we consider the formal matrix model.

$\bullet$ In the 3rd section we perform the sum over filling fractions, and we get $\Theta$-functions.

$\bullet$ In the 4rth section we discuss the link with degenerate Riemann surfaces and holomorphic anomaly equations.

$\bullet$ In the 5th section we discuss the background independence problem.

$\bullet$ Section 6 is the conclusion.

\subsection{Introductory example: 1 matrix model}

\subsubsection{Paths and homology basis}

Consider a polynomial potential $V(x)$, of degree $d+1=\deg V$, with complex coefficients.
There are many different paths $\gamma$ such that the integral
\beq
\int_{\gamma} dx\, \ee{-V(x)}
\eeq
is absolutely convergent.
These are the paths which go to $\infty$ in a sector where $\Re V>0$, or more precisely, the paths which connect two such sectors (see \cite{Marcopath} for a discussion on that, or \cite{BEHRH}. Those considerations can be easily extended to any $V$ such that $V'$ is a rational fraction).

\medskip
{\bf Example:} quartic potential $V(x)=x^4$, we have $\deg V=4$, i.e. there are $d=3$ independent paths, for example we choose:
{\mbox{\epsfxsize=4.truecm\epsfbox{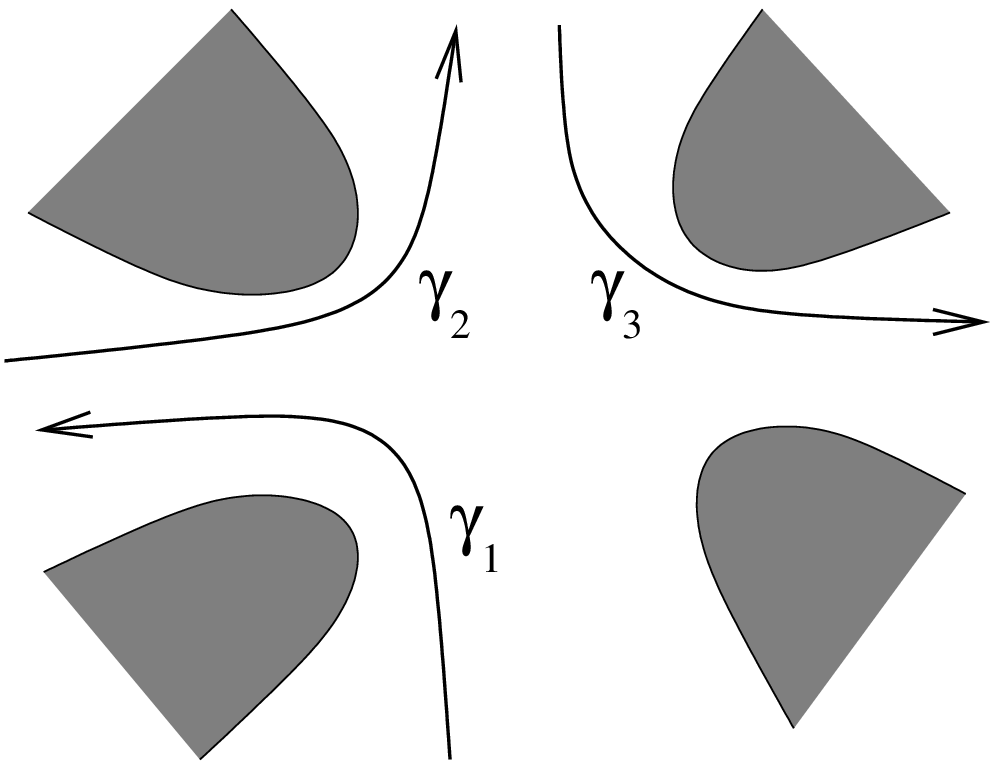}}}.
In this example, we have ${\mathbb R} = \gamma_2+\gamma_3$.

\medskip

In fact, there are $d=\deg V'$ homologically independent such paths. Let us choose a basis:
\beq
\gamma_1,\dots,\gamma_d
\eeq
This means any (unbounded) path on which the integral $\int_{\gamma} dx\, \ee{-V(x)}$ is well defined, can be decomposed on the basis:
\beq
\gamma=\sum_{i=1}^d c_i\, \gamma_i
\eeq
By definition:
\beq
\int_{\gamma} dx\, \ee{-V(x)} = \sum_{i=1}^d c_i\,\int_{\gamma_i} dx\, \ee{-V(x)}
\eeq
In this definition, the $c_i$'s can be arbitrary complex numbers, they don't need to be integers.

However, if $\gamma$ is itself a path, the $c_i$'s can take only the values $+1,-1,$ or $0$.

If the $c_i$'s are not integers, we say that $\gamma=\sum_i c_i \gamma_i$ is a {\bf generalized path}.

\subsubsection{Matrix model on a generalized path}

Let $\gamma$ be a generalized path.
We define the set of {\bf Normal matrices on $\gamma$}:

\beq
H_n(\gamma) = \big\{ M=U\, {\rm diag}(x_1,\dots,x_n)\, U^\dagger \,\, / \,\, U\in U(n)\, , \,\, \forall i\,\, x_i\in \gamma \big\}
\eeq
equipped with the measure:
\beq
dM = \Delta(x)^2\,\,\, dU\,\, dx_1\,\dots\,dx_n
\virg
\Delta(x)=\prod_{i<j}(x_j-x_i)
\eeq
where $dU$ is the Haar measure on $U(n)$, and $\Delta(x)$ is the Vandermonde determinant, and $dx$ is the curviline measure along the path (if $\gamma=x(s)\, , \,\, s\in {\mathbb R}$, is a parametrization of the path we have $dx=x'(s)\, ds$).

\medskip
{\bf Remark:}
$H_n({\mathbb R}) = H_n$ is the set of hermitian matrices, with the usual $U(n)$ invariant measure.

\medskip
{\bf Remark:}
In general $H_n(\gamma)$ is not a group, for instance the sum of two matrices in $H_n(\gamma)$ is not in $H_n(\gamma)$, and the product by a scalar is not either.
Also, the "measure" $dM$ is not positive, in fact it is complex.

\bigskip

The matrix integral on $H_n(\gamma)$ is defined as follows:
\beq\label{defintconv}
\hat{Z}(\gamma) = {1\over n!}\,\int_{H_n(\gamma)} dM\,\, \ee{-N \Tr V(M)} =  {1\over n!}\,\int_{\gamma^n} dx_1\dots dx_n\,\, \Delta(x)^2 \,\, \prod_i \ee{-N V(x_i)}
\eeq
or in other words:
\beq\label{ZgammasumZff}
\hat{Z}(\gamma) = \sum_{n_1+\dots+n_d=n} \,c_1^{n_1}\dots c_d^{n_d}  \,\,\, Z(n_1/N,\dots, n_d/N)
\eeq
where we have defined:
\beq\label{defZfixfilling}
Z(n_1/N,\dots, n_d/N) = {1\over n_1!\dots n_d !}\,\,\, \int_{\gamma_1^{n_1}\times\dots \times \gamma_d^{n_d}} dx_1\dots dx_n\,\, \Delta(x)^2 \,\, \prod_i \ee{-N V(x_i)}
\eeq

\subsubsection{Assumption: topological expansion}

First, let us assume that only $\genus+1\leq d$ of the $c_i$'s are non-vanishing.
We write:
\beq\label{coefci}
\forall i=1,\dots,\genus\, , \quad c_i = \ee{2i\pi \nu_i}
\virg
c_{\genus+1}=1
\virg
\forall i=\genus+2,\dots,d\, , \quad c_i = 0
\eeq
If $\gamma$ is a path, the $c_i$'s take the values $\pm 1$, and thus $\nu_i=0$ or $1/2$.
The vector $(\nu_1,\dots,\nu_{\genus})$ is going to be considered a charcateristic in a genus $\genus$ Jacobian.
Also, up to reverting the orientation of $\gamma_i$, we can always assume that if $\gamma$ is a path, 
\beq\label{hypnuzero}
\gamma={\rm path}\quad \Rightarrow\quad
\forall i=1,\dots,\genus+1,\,\,\, c_i=1
\virg
\Rightarrow \nu=0
\eeq

\medskip
{\bf Hypothesis:}

Our working hypothesis is that the basis paths $\gamma_1,\dots,\gamma_{\genus+1}$ have been chosen so that each $Z(n_1/N,\dots, n_d/N)$ admits a large $N$ topological expansion:
\beq
\ln{(Z(\epsilon_1,\dots,\epsilon_{d-1}))} \sim F(\bfeps)=\sum_{h=0}^\infty N^{2-2h} F^{(h)}(\bfeps)
\eeq

It is conjectured that given a (generic) potential $V$, and a generalized path $\gamma$, such a "good" basis always exists (may be not unique). In fact, for the 1-matrix model with polynomial potential, this can be proved a posteriori from the asymptotics of M. Bertola \cite{Bertasympt, BMoasymp}.
But for more general cases, it is only a conjecture, for instance for the 2-matrix model.

Now, let us explain where this hypothesis comes from, and what heuristic arguments support it.

\subsubsection{Loop equations and Virasoro constraints}

It is well known that any integral defined in eq.\ref{defZfixfilling}, satisfies an infinite set of linear equations, sometimes called "loop equations" \cite{ZJDFG}, or Virasoro constraints, or Schwinger--Dyson equations, or Euler--Lagrange equations, and which just come from integration by parts:
\beq
\forall k\geq -1\, , \qquad {\cal V}_k.Z = 0
\eeq
\beq
{\cal V}_k = \sum_{j=1}^{\deg V} j t_j {\partial \over \partial t_{k+j}} + {1\over N^2}\, \sum_{j=0}^k {\partial \over \partial t_{j}} \,{\partial \over \partial t_{k-j}}
\virg
V(x) = \sum_j t_j x^j
\eeq
They satisfy Virasoro algebra:
\beq
[{\cal V}_k,{\cal V}_j]=(k-j){\cal V}_{k+j}
\eeq

{\bf Remark:}
It is important to notice that, since integration by parts is independent of the integration paths (as long as there is no boundary term), both $\hat{Z}(\gamma)$ and any $Z(n_1/N,\dots, n_d/N),\,\, \forall n_i$, satisfy the same set of loop equations.

\subsubsection{Formal matrix models and combinatorics of maps}

Formal matrix integrals are defined as formal generating functions for enumerating discrete surfaces (also called "maps", i.e. topological graphs embedded on a Riemann surface, such that each face is a disc) of given topology.
Basically, $F_g$ is the generating function for counting "maps" of genus $g$.
The generating series:
\beq
\ln{Z_{\rm formal}} = \sum_{g=0}^\infty N^{2-2g} F_g
\eeq
needs not be convergent, and in fact it is never convergent if the weights for "maps" are positive.
It is merely a formal series, whose only role is to encode the $F_g$'s.

\medskip
The formal matrix integrals satisfy the same loop equations, i.e. Virasoro constraints as $\hat{Z}(\gamma)$ and $Z(n_1/N,\dots, n_d/N)$ (see \cite{ZJDFG}).
In the context of combinatorics of maps, loop equations are known as Tutte's equations \cite{tutte2}, and were first obtained by counting "maps" recursively (removing one edge at each step).

\medskip
The $F_g$'s of formal matrix integrals have all been computed: $F_0$ has been known for a long time, then $F_1$ \cite{ChekhovF1}, and all the $F_g$'s with $g\geq 2$ were computed recently in \cite{eynch, EOFg}.

\medskip
In fact, it was proved in \cite{eynch, EOFg}, that any solution of loop equations which has a topological large $N$ expansion of the form:
\beq
\ln{Z} = \sum_{g=0}^\infty N^{2-2g} F_g
\eeq
can be obtained by the symplectic invariants of \cite{EOFg}, i.e. they are encoded by a spectral curve.

\subsubsection{Spectral curves}

Both the convergent matrix integral, and the formal matrix integral are associated to an (algebraic) spectral curve of the form:
\beq
y^2 = V'(x)^2 - {4\over N} \left< \Tr {V'(x)-V'(M)\over x-M}\right>
\eeq

$\bullet$ For the convergent matrix integral $\hat{Z}$ defined in eq.\ref{defintconv}, the average $<.>$ is taken with respect to the measure $dM\,\, \ee{-N\Tr V(M)}$. The notion of a spectral curve, comes from the orthogonal polynomials method of Dyson-Mehta \cite{Mehta}, combined with the theory of integrable systems \cite{Babook}. The orthogonal polynomials satisfy an integrable differential equation of the form  $\,\vec\psi' = {\cal D}(x)\, \vec\psi$, where ${\cal D}(x)$ is a $2\times 2$ matrix with polynomial coefficients, and the spectral curve is by definition the set of eigenvalues of ${\cal D}$ (Jimbo-Miwa-Ueno \cite{Miwa}), i.e.:
\beq
y^2 = {1\over 2}\Tr {\cal D}(x)^2
\eeq
It was proved \cite{BEHiso} that:
\beq
{1\over 2}\Tr {\cal D}(x)^2  = V'(x)^2 - {4\over N} \left< \Tr {V'(x)-V'(M)\over x-M}\right>
\eeq

$\bullet$ For the formal matrix model, and more generally, for an arbitrary solution of the Virasoro constraints which has a topological expansion, the average $<.>$ has a formal meaning, and can be defined from the Virasoro generators ${\cal V}_k$. It is not the purpose of this article to explain where it comes from (see \cite{eynform, ZJDFG}), and the spectral curve is the algebraic equation satisfied by the "disc amplitude", i.e. generating function for counting planar "maps" with one boundary (i.e. having the topology of a discs), and it can be proved that it satisfies:
\beq
y^2 = V'(x)^2 - 4 P(x)
\eeq
where $P(x)$ is a polynomial of degree $d-1=\deg V''$, and with the same leading coefficient as $V'$.
\beq
P(x) = (d+1)\, t_{d+1}\, x^{d-1} + \sum_{k=0}^{d-2}\, P_k\, x^k
\eeq
The coefficients $P_k$, are the conserved quantities in the context of integrable systems \cite{Babook}, whereas the coefficients of $V'$ are called the Casimirs.
The coefficients $P_k$ are in 1-1 correspondance with the so-called "action variables":
\beq
\epsilon_i = {1\over 2i\pi}\, \oint _{\acycle_i} y dx
\virg
i=1,\dots,d-1
\eeq
Here in the random matrix context, the $\epsilon_i$'s are called {\bf filling fractions}.

\subsubsection{Symplectic invariants}

In \cite{EOFg}, it was proved, that given a spectral curve 
\beq
{\cal E}(x,y)=0
\eeq
(here ${\cal E}(x,y)=y^2 - V'(x)^2 + 4 P(x)$, i.e. in other words, given a potential $V(x)=\sum_{k=1}^{d+1} t_k x^k$ and a polynomial $P(x) = (d+1)\, t_{d+1}\, x^{d-1} + \sum_{k=0}^{d-2}\, P_k\, x^k$, or in other words given $V'$ and the filling fractions $\epsilon_i$'s), it is possible to define an infinite sequence:
\beq
F_g({\cal E})\virg g=0,\dots,\infty
\eeq
such that:
\beq
\tau({\cal E}) = \exp{\sum_{g=0}^\infty N^{2-2g}F_g({\cal E})}
\eeq
is a solution of loop equations.

The $F_g({\cal E})$ were constructed in \cite{EOFg} for any spectral curve ${\cal E}(x,y)=0$, and they have many interesting properties, for instance they are invariant under {\bf symplectic} deformations of the spectral curve, and $\tau({\cal E})$ is the $\tau$-function of an {\bf integrable} hierarchy.
Their modular properties were also studied in \cite{EOFg} and further clarified in \cite{eynHolan}, and they happen to be deeply related to the so-called {\bf Holomorphic anomaly equation} first found in string theory \cite{BCOV, abk}, and which relate the non-holomorphic part of the generating function for counting Riemann-surfaces to the contribution of degenerate Riemann surfaces (nodal surfaces).
This will play a role below.

\medskip
Also, in \cite{EOFg}, were defined the correlators:
\beq
W_k^{(g)}(z_1,\dots,z_k)
\virg g=0,\dots,\infty\,\, , \,\,
k=0,\dots,\infty
\virg
(\,\,W_0^{(g)}=F_g\,\,)
\eeq
which are multilinear symmetric meromorphic differential forms on the spectral curve.
They also have many interesting properties, in particular they can be used to compute derivatives of the $F_g$'s with any parameter on which ${\cal E}$ may depend.
For instance derivatives with respect to filling fractions are:
\beq\label{derWngff}
{\partial \over \partial \epsilon_j}\, W_k^{(g)}(z_1,\dots,z_k) = \oint_{\bcycle_j} W_{k+1}^{(g)}(z_1,\dots,z_k,z_{k+1}) 
\eeq
(where $\tau$ is the Riemann matrix of periods of the spectral curve, and $\acycle_i\cap \bcycle_j=\delta_{i,j}$ is a symplectic basis of non contractible cycles, see \cite{Farkas, Fay} for algebraic geometry).

\subsubsection{Heuristic support to the conjecture}

The conjecture is supported by the following facts:

$\bullet$ Both the convergent matrix integral  $Z({n_1\over N},\dots,{n_{\genus+1}\over N},0,\dots,0)$ defined in eq.\ref{defZfixfilling}, and the formal matrix integral $Z_{\rm formal}(\epsilon_1,\dots,\epsilon_{\genus})$ satisfy the same loop equations.

$\bullet$ Since loop equations are linear, the space of solutions is a vector space.

$\bullet$ For given $V'$, both the convergent integral $Z({n_1\over N},\dots,{n_{\genus+1}\over N},0,\dots,0)$, and the formal $Z_{\rm formal}(\epsilon_1,\dots,\epsilon_{\genus})$ are specified by the same number of parameters, i.e. $\genus$ parameters (indeed $n_1+\dots+n_{\genus+1}=n$, so that only $\genus$ of them are independent).

\bigskip
Those observations support the idea that there exists a good basis of the vector space of solutions, such that each basis function is at the same time formal and convergent, i.e. there exists a set of basis paths $\gamma_i$, such that $Z({n_1\over N},\dots,{n_{\genus+1}\over N},0,\dots,0)$ admits a topological expansion.

 \bigskip
 We do not prove this conjecture in this article, but we take it as an asumption.

\subsection{Generalization 2-Matrix model}

All this can be extended to a larger context, for instance the 2-matrix model, or the chain of matrices, or the matrix model coupled to an external field.

\bigskip

{\bf 2 matrix model}

Consider 2 polynomial potentials $V_1$ and $V_2$, such that $\deg V_1=d_1+1, \deg V_2=d_2+1$.
There are $d_1\times d_2$ independent paths on ${\mathbb  C}\times{\mathbb C}$ on which the following integral is absolutely convergent:
\beq
\int\int_\gamma\,\, dx\, dy\,\, \ee{-V_1(x)-V_2(y)+xy}
\virg
\gamma=\sum_{i=1}^{d_1 d_2}\,\, c_i \gamma_i
\eeq
where each $\gamma_i$ is a product of a path in the $x-$plane and a path in the $y-$plane.

Then, similarly to the 1-matrix case, we can also define a matrix integral on a generalized path (see \cite{eynhabilit}):
\beq
\hat{Z}(\gamma) = \int_{H_n\times H_n(\gamma)}\,\, dM_1\, dM_2\,\,\ee{-N\Tr(V_1(M_1)+V_2(M_2)-M_1 M_2)}
\eeq
which satisfies:
\beq
\hat{Z}(\gamma) = \sum_{n_1+\dots+n_d=n} \,c_1^{n_1}\dots c_d^{n_d}  \,\,\, Z(n_1/N,\dots, n_d/N)
\eeq
where we have defined:
\bea
Z(n_1/N,\dots, n_d/N) 
&=& {1\over n_1!\dots n_d !}\,\,\, \int_{\gamma_1^{n_1}\times\dots \times \gamma_d^{n_d}} dx_1\wedge dy_1\,\dots dx_n\wedge dy_n\,\, \cr
&& \qquad \Delta(x)\Delta(y) \,\, \prod_i \ee{-N (V_1(x_i)+V_2(y_i)-x_i y_i)}
\eea

The 2-matrix model generalized integral satisfies loop equations (which form a W-algebra instead of Virasoro), which also come from integration by parts, and are independent of the path.
In particular, each $Z(n_1/N,\dots, n_d/N)$ satisfies the same loop equations.

\medskip

There is also a formal 2-matrix model, which was introduced as a generating function for bi-colored discrete surfaces, it was called the "Ising model on a random lattice" \cite{KazIsing}.
Almost by definition, the formal 2-matrix model has a topological expansion:
\beq
\ln{Z} = \sum_g N^{2-2g} F_g
\eeq
The formal 2-matrix model satisfies the same loop equations as the convergent one, and the solution of loop equations was found in \cite{EOloopeq,CEOloopeq,EOFg}, and it was found that the $F_g$'s are again the symplectic invariants of \cite{EOFg}.

\bigskip
{\bf matrix model with external field}

The same features also hold for the matrix models in an external field.
The famous example is the Kontsevich integral \cite{kontsevitch}, also called "matrix Airy function":
\beq
Z_{\rm Kontsevich} = \int dM\,\, \ee{-N\Tr {M^3\over 3}-M\L}
\eeq
whose topological expansion is the combinatorics generating function computing intersection numbers.

\bigskip
{\bf Summary}

In all cases, there is a convergent matrix model defined on generalized paths, and there is a formal matrix model which computes the combinatorics of some graphs.
Both the convergent and formal model obey the same set of loop equations. 

The formal model has a topological expansion
\beq
\ln{Z} = \sum_g N^{2-2g} F_g
\eeq
where the $F_g$'s are the symplectic invariants of \cite{EOFg}, computed for some algebraic spectral curve ${\cal E}(x,y)=0$. And in all cases the dimension of the homology basis of paths on which the integral is absolutely convergent, is the same as the genus $\genus$ of the spectral curve, i.e. the number of filling fractions:
\beq
\gamma=\sum_{i=1}^{\genus+1} c_i \gamma_i
\quad \Leftrightarrow \quad
\epsilon_i = {1\over 2i\pi}\, \oint _{\acycle_i} y dx
\virg
i=1,\dots,\genus
\eeq

In all those cases, the method we describe below should work.

\section{Formal matrix model}

Now, assume that $Z(\epsilon_1,\dots,\epsilon_{d-1})$ has a topological asymptotic expansion:
\beq
\ln{(Z(\epsilon_1,\dots,\epsilon_{d-1}))} = F(\bfeps)=\sum_{h=0}^\infty N^{2-2h} F_h(\bfeps)
\eeq
Each $F_h$ must then be a solution of formal loop equations, and therefore it is given by the formulae of \cite{EOFg}, and therefore each $F_h$ is analytical in the $\epsilon_i$'s.

\medskip

Then, we choose arbitrarily a "prefered" filling fraction $\bfeps^*$, and perform the 
Taylor expansion:
\beq\label{FhTaylexp}
F_h(\bfeps)=\sum_{k=0}^\infty {1\over k!}\,F_h^{(k)} (\bfeps-\bfeps^*)^k
\virg
F_h^{(k)}  = {\d^k F_h \over \d\bfeps^k}(\bfeps^*)
\eeq
{\bf Remark:}
We don't write the indices for readability, but $F_h^{(k)}$ is a tensor. 
For readability we write the formulae as if there were only one variable $\epsilon$, i.e. $\genus=1$, but in fact we mean:
\beq
F_h(\bfeps)=\sum_{k=0}^\infty {1\over k!}\, \sum_{i_1,\dots,i_k}\,{F_h^{(k)}}_{i_1,\dots,i_k}\,\, \prod_{j=1}^k (\bfeps-\bfeps^*)_{i_j}
\virg
{F_h^{(k)}}_{i_1,\dots,i_k}  = {\d^k F_h \over \d\bfeps_{i_1}\dots \d\bfeps_{i_k}}(\bfeps^*)
\eeq
but for simplicity we shall write eq.\ref{FhTaylexp}, and we leave to the reader to restore the indices if needed.

\medskip
The derivatives of $F_g$, are given by eq.\ref{derWngff} (see \cite{EOFg}):
\beq\label{derFgff}
{F_h^{(k)}}_{i_1,\dots,i_k}= {\partial^l \over \partial \epsilon_{i_1}\dots \partial \epsilon_{i_k}}\, F_h = \oint_{\bcycle_{i_1}}\dots \oint_{\bcycle_{i_k}} W_{k}^{(h)}(z_1,\dots,z_k) 
\eeq

\medskip

In particular, it is well known (see \cite{EOFg}), that
\beq
F'_0 = \oint_{\bcycle} ydx
\eeq
and ${1\over 2i\pi}F_0''= \tau$ is the Riemann matrix of periods (see \cite{Farkas, Fay} for introduction to algebraic geometry) of the specral curve ${\cal E}$, i.e.
\beq
{1\over 2i\pi}\,{\partial^2 F_0\over \partial \epsilon_i \partial \epsilon_j} = \tau_{i,j}=\tau_{j,i} = \oint_{\bcycle_i} du_j
\eeq
where $du_j$ is the normalized basis of holomorphic differentials \cite{Farkas, Fay} on ${\cal E}$:
\beq
\oint_{\acycle_i} du_j = \delta_{i,j}
\eeq

And thus we have (formally):
\bea\label{TaylexpZ}
Z(\bfeps)
&=& Z(\bfeps^*)\,\ee{i\pi N^2(\bfeps-\bfeps^*)\tau(\bfeps-\bfeps^*)}\,\ee{2i\pi N^2\bfzeta(\bfeps-\bfeps^*)}  
 \sum_{k} \sum_{l_i}\sum_{h_i} \cr
 && \qquad {N^{\sum_i (2-2h_i)}\over k! l_1!\,\dots\, l_k!}\,\,\, F_{h_1}^{(l_1)}\dots F_{h_k}^{(l_k)}  \,\, (\bfeps-\bfeps^*)^{\sum l_i}
\eea
where we the sum carries only on $l_i\geq 1$ and $2-2h_i-l_i<0$ for all $i$.

One should notice that the exponential is now at most quadratic in the $\epsilon$'s.

\section{Oscillations}

Now we are going to perform the sum of eq.\ref{ZgammasumZff}:
\beq
\hat{Z}(\gamma) = \sum_{n_1+\dots+n_{\genus+1}=n} \,c_1^{n_1}\dots c_{\genus+1}^{n_{\genus+1}}  \,\,\, Z(n_1/N,\dots, n_d/N)
\eeq
where
\beq
\gamma = \sum_i c_i \gamma_i
\virg
c_i = \ee{2i\pi\, \nu_i}
\eeq
Since the filling fractions $\epsilon_i = {n_i\over N}$ take integer values (up to a $1/N$ factor), we have to perform a sum of exponentials of square of integers.
Such sums are called {\bf theta functions}. They play a key role in algebraic geometry.
Let us recall a few properties \cite{Farkas, Fay}.

\subsection{Theta functions}

We define the $\Theta$-function:
\beq
\Theta(u,t) = \sum_{\bfn\in {\mathbb Z}^\genus} \ee{(n-N\bfeps^*)u}\,\,\ee{(n-N\bfeps^*)t(n-N\bfeps^*)}\,\,\ee{2i\pi\, \bfn\nu}
\eeq
It clearly satisfies:
\beq\label{eqdifTheta}
{\partial \Theta\over \partial t}
= {\partial^2 \Theta\over \partial u^2}
\eeq
It is related to the usual Jacobi-theta function: 
\beq
\Theta(u,t) =
\theta_{-N\bfeps^*,\nu}({u\over 2 i\pi},{t\over i\pi})\,\, \ee{2i\pi \nu N\bfeps^*}
\eeq
where $(-N\bfeps^*,\nu)$ is called the characteristics. The Jacobi theta function with characteristics $(a,b)$ is defined by:
\beq
\theta_{a,b}(u,\tau) = \sum_n \ee{2i\pi(n+a)(u+b)}\,\ee{i\pi(n+a)\tau(n+a)}
=\theta_{0,0}(u+b+\tau a,\tau)\,\, \ee{i\pi a\tau a} \,\, \ee{2i\pi a (u+b)}
\eeq
It takes a phase after translation along an integer lattice period $n+\tau m$:
\beq
\theta_{a,b}(u+n+\tau m,\tau) = \ee{2i\pi (an-mb)}\,\theta_{a,b}(u,\tau)\,\,\ee{-2i\pi m u}\,\ee{-i\pi m\tau m}
\eeq

\subsection{Convergent matrix model}

We thus have:
\bea
\hat{Z}(\gamma)
&\sim &  \sum_{\bfn} \ee{2i\pi\, \bfn\nu} \, Z_{\rm formal}(\bfn/N)   \cr
&\sim& \sum_{\bfn} c_1^{n_1}\dots c_{\genus}^{n_{\genus}}\,\, Z(n_1/N,\dots,n_\genus/N,0,\dots,0)
\eea
The sum carries on integers $n_i\geq 0$ and $\sum_i n_i = n$. 
Therefore $n_{\genus+1}=n-\sum_{i=1}^\genus n_i$ is not independent from the others.
Another remark, is that in that sum we expect that only the vicinity of some extremal $n_i$ will dominate the sum, and that values of the $n_i$'s far from the extremum should give an exponentially small contribution.
That asumption allows to extend the sum to $n_i\in {\mathbb Z}$.

Then, we use the Taylor expansion of eq.\ref{TaylexpZ}, and we find (again we use tensorial notations):
\bea\label{TaylexpZsum}
\hat{Z}(\gamma)
&\sim& Z(\bfeps^*)\, \sum_{\bfn\in {\mathbb Z}^\genus}\, \ee{2i\pi \sum_i \nu_i n_i}\,\,\ee{i\pi (\bfn-N\bfeps^*)\tau(\bfn-N\bfeps^*)}\,\ee{2i\pi N\bfzeta(\bfn-N\bfeps^*)}  \cr
&& \qquad  \sum_{k} \sum_{l_i>0}\sum_{h_i>1-{l_i\over 2}} {N^{\sum_i (2-2h_i-l_i)}\over k! l_1!\,\dots\, l_k!} F_{h_1}^{(l_1)}\dots F_{h_k}^{(l_k)}  \,\, (\bfn-N\bfeps^*)^{\sum l_i} \cr
\eea
where we recognize the $\Theta$-function and its derivatives
\beq\label{exphatZTheta}
\encadremath{
\hat{Z}(\gamma)
\sim   Z(\bfeps^*)\,  \sum_{k} \sum_{l_i>0}\sum_{h_i>1-{l_i\over 2}} {N^{\sum_i (2-2h_i-l_i)}\over k! l_1!\,\dots\, l_k!} F_{h_1}^{(l_1)}\dots F_{h_k}^{(l_k)}  \,\, \left.{\partial^{\sum l_i} \Theta\over \partial u^{\sum l_i}}\right|_{u=N F_0',t=i\pi\tau}
}\eeq
This formula is the main result presented in this article.

For instance the first few terms in powers of $N^{-1}$ are:
\bea
\hat{Z}(\gamma)
&\sim&  Z(\bfeps^*)\, \Big( \Theta + {1\over N} (F_1' \Theta' + {F_0'''\over 6}  \Theta''' ) \cr
&&  + {1\over N^2} ({F_1''\over 2} \Theta'' + {(F_1')^2\over 2}  \Theta'' + {F_0''''\over 24} \Theta^{(4)}  + {F_1'\, F_0'''\over 6}  \Theta^{(4)} + {(F_0''')^2\over 72}  \Theta^{(6)} ) \cr
&&  +\dots \Big)
\cr
\eea

\subsection{Resummation}

The expansion of formula .\ref{exphatZTheta} can be resummed into a single $\Theta$-function.
We want to write it as:
\beq\label{Thetaresummed}
\hat{Z}(\gamma) =  Z(\bfeps^*)\,  \Theta(u,t)
\eeq
where
\beq
u= N F_0' + \sum_{h=1}^\infty N^{1-2h} u^{(h)} 
\virg
t = i\pi \tau + \sum_{h=1}^\infty N^{-2h} t^{(h)}
\eeq

For instance, one easily finds the first orders:
\beq
u^{(1)} = F_1' + {F_0'''\over 6}\,{\Theta'''(u^{(0)},i\pi\tau)\over \Theta'(u^{(0)},i\pi\tau)}
\eeq
\beq
t^{(1)} = {F_1''\over 2} + {F_0''''\over 24}\,{\Theta''''(u^{(0)},i\pi\tau)\over \Theta''(u^{(0)},i\pi\tau)}+
{F_1'\, F_0'''\over 6} \left( {\Theta^{(4)}\over \Theta''}-{\Theta^{'''}\over \Theta'}\right) + {(F_0''')^2\over 72}  \left({\Theta^{(6)}\over \Theta''} - {\Theta'''^2\over \Theta'^2}  \right)
\eeq

\medskip
The Taylor expansion of eq.\ref{Thetaresummed} reads (and we use eq.\ref{eqdifTheta}):
\bea
\hat{Z}(\gamma) 
&=& Z(\bfeps^*)\,\, \Theta(NF_0'+{1\over N}u^{(1)}+\dots, i\pi \tau + {1\over N^2} t^{(1)}+\dots) \cr
&=&  Z(\bfeps^*)\,  \sum_{m,n}  {(m+n)!\over m! n!}\,\,(u-u^{(0)})^m (t-t^{(0)})^n  {\partial^{m+2n}\over \partial u^{m+2n}} \Theta(u^{(0)},t^{(0)})  \cr
&=&  Z(\bfeps^*)\,  \sum_{m,n} 
\sum_{k_1,\dots,k_m}\sum_{j_1,\dots,j_n}  {(m+n)!\,N^{m-2\sum k_i - 2\sum j_i }\over m!\, n!}\, \,\, \cr
&& \qquad u^{(k_1)}\dots u^{(k_m)} \,\,t^{(j_1)}\dots t^{(j_n)}  {\partial^{m+2n}\over \partial u^{m+2n}} \Theta(u^{(0)},t^{(0)})  
\eea
and now we identify the powers of $N$ with equation.\ref{exphatZTheta}.
For any given $p>0$, we must have:
\bea\label{identThetaFgresum}
&& \sum_{k_1,\dots,k_m}\sum_{j_1,\dots,j_n}  {(m+n)!\over m! n!}\,\, u^{(k_1)}\dots u^{(k_m)} \,\,t^{(j_1)}\dots t^{(j_n)}  \,\,\partial_u^{m+2n} \Theta(u^{(0)},t^{(0)})  \cr
&=&
  \sum_{r} \sum_{l_i}\sum_{h_i} {1\over r! l_1!\,\dots\, l_r!} F_{h_1}^{(l_1)}\dots F_{h_r}^{(l_r)}  \,\, \partial_u^{\sum l_i} \Theta(u^{(0)},t^{(0)})
\eea
where in the first sum, the indices are such that 
\beq
p= 2\sum_{i=1}^m k_i + 2 \sum_{i=1}^n j_i - m\virg k_i>0,\,\, j_i>0
\eeq
and in the second sum
\beq
p=  \sum_{i=1}^r (2h_i +  l_i -2) 
\virg l_i>0,\,\, 2-2h_i-l_i<0
\eeq
This equation defines $u^{(k)}$ and $t^{(l)}$ recursively in a unique way.

Indeed, assume that we have already computed $u^{(1)},\dots,u^{(q-1)}$ and $t^{(1)},\dots, t^{(q-1)}$.
Choose $p=2q-1$ in eq.\ref{identThetaFgresum}:
\bea
&& u^{(q)}\,\Theta'(u^{(0)},t^{(0)}) \cr
&=& \sum_{r} \sum_{l_i}\sum_{h_i} {1\over r! l_1!\,\dots\, l_r!} F_{h_1}^{(l_1)}\dots F_{h_r}^{(l_r)}  \,\, \partial_u^{\sum l_i} \Theta(u^{(0)},t^{(0)}) \cr
&&  - \sum_{k_1,\dots,k_m}\sum_{j_1,\dots,j_n}  {(m+n)!\over m! n!}\,\, u^{(k_1)}\dots u^{(k_m)} \,\,t^{(j_1)}\dots t^{(j_n)}  \,\,\partial_u^{m+2n} \Theta(u^{(0)},t^{(0)}) \cr
\eea
where in the first sum we have $2q-1=\sum_{i=1}^r (2h_i+l_i-2), l_i>0, 2-2h_i-l_i<0$, and in the second sum we have $2q-1= 2\sum_{i=1}^m k_i+2\sum_{i=1}^n j_i - m$, which implies $k_i<q$ and $j_i<q$, i.e. all the terms in the RHS are known from the recursion hypothesis.
We have thus determined $u^{(q)}$.
Then, let $p=2q$, we have:
\bea
&& t^{(q)}\,\Theta''(u^{(0)},t^{(0)}) \cr
&=& \sum_{r} \sum_{l_i}\sum_{h_i} {1\over r! l_1!\,\dots\, l_r!} F_{h_1}^{(l_1)}\dots F_{h_r}^{(l_r)}  \,\, \partial_u^{\sum l_i} \Theta(u^{(0)},t^{(0)}) \cr
&& \quad - \sum_{k_1,\dots,k_m}\sum_{j_1,\dots,j_n}  {(m+n)!\over m! n!}\,\, u^{(k_1)}\dots u^{(k_m)} \,\,t^{(j_1)}\dots t^{(j_n)}  \,\,\partial_u^{m+2n} \Theta(u^{(0)},t^{(0)}) \cr
\eea
where in the first sum we have $2q=\sum_{i=1}^r (2h_i+l_i-2), l_i>0, 2-2h_i-l_i<0$, and in the second sum we have $(m,n)\neq (0,1)$, $2q= 2\sum_{i=1}^m k_i+2\sum_{i=1}^n j_i - m$, which implies $k_i\leq q$ and $j_i<q$, i.e. all the terms in the RHS are known from the recursion hypothesis.
We have thus determined $t^{(q)}$.

\bigskip

Therefore we have:
\beq\label{ZThetaresummed}
\encadremath{
\hat{Z}(\gamma) 
= Z(\bfeps^*)\,\, \Theta\left(NF_0'+\sum_k N^{1-2k} u^{(k)}, i\pi \tau + \sum_{j} N^{-2j} t^{(j)}\right) 
}\eeq

It would be interesting to understand how this formula matches the tau-function obtained from integrability properties \cite{Babook}.

\section{Holomorphic anomaly equations}

One may observe that all the terms with even powers of $N$ in formula eq.\ref{exphatZTheta} have already appeared in another context, in topological string theory \cite{mmhouches}, and more precisely the so called "holomorphic anomaly equations" \cite{BCOV}.

Holomorphic anomaly equations are about modular invariance versus holomorphicity.

Let us briefly sketch the idea.
String theory partition functions represent "integrals" over the set of all Riemann surfaces with some conformal invariant weight. In other words, they are integrals over moduli spaces of Riemann surfaces of given topology, and topological strings are integrals with a topological weight, they compute intersection numbers (see \cite{vonk, mmhouches} for introduction to topological strings).

Moduli spaces can be compactified by adding their "boundaries", which correspond to degenerate Riemann surfaces (for instance when a non contractible cycle gets pinched). The integrals have thus boundary terms, which can be represented by $\delta$-functions, and $\delta$-functions are not holomorphic.
In other words, string theory partition functions contain non-holomorphic terms which count degenerate Riemann surfaces.

On the other hand, if one decides to integrate only on non-degenerate surfaces, one gets holomorphic patition functions, but not modular invariant, because the boundaries of the moduli spaces are associated to a choice of pinched cycles.
Modular invariant means independent of a choice of cycles.

To summarize, the holomorphic partition function is obtained after a choice of boundaries, i.e. a choice of a symplectic basis of non contractible cycles $\acycle_i\cap \bcycle_j=\delta_{i,j}$, and cannot be modular invariant.
The modular invariance is restored by adding the boundaries, but this breaks holomorphicity.

There is thus a relationship between holomorphicity and modular invariance.

Let $F_g$ be the partition function corresponding to the moduli space of non-degenerate  Riemann surfaces of genus $g$, i.e. $F_g$ is holomorphic but not modular invariant (it assumes a choice of a basis of cycles $\acycle_i$, $\bcycle_i$, $i=1,\dots,g$), and let $\hat{F}_g$ be the partition function including degenerate surfaces, i.e. non holomorphic but modular invariant.
The holomorphic anomaly equation discovered by \cite{BCOV}, states that:
\beq
\overline{\partial} \hat{F}_g = \,{1\over 2}\,{\overline\partial \kappa}\,\,\left( \hat{F}_{g-1}'' + \sum_{h=1}^{g-1} \hat{F}_h' \hat{F}_{g-h}' \right)
\eeq
where $\kappa$ is the Zamolodchikov K\"ahler metric symmetric matrix:
\beq
\kappa=(\overline{\tau}-\tau)^{-1}
\eeq

It was found in \cite{BCOV, abk, eynHolan} that:
\bea\label{eqZhatkappa}
\hat{Z} 
&=& \ee{\sum_g N^{2-2g} \hat{F}_g} \cr
&=& \ee{\sum_gN^{2-2g} F_g}\, \sum_l\sum_{k} \sum_{l_i>0}\sum_{h_i>1-{l_i\over 2}} {N^{\sum_i (2-2h_i-l_i)}\over k! l_1!\,\dots\, l_k!}\cr
&& \qquad \quad F_{h_1}^{(l_1)}\dots F_{h_k}^{(l_k)}  \,\, (2l-1)!!\,\,\kappa^l\,\,\,\delta_{2l-\sum l_i} \cr
\eea
Remember that we use tensorial notations, and 
\beq
(2l-1)!!\,\,\kappa^l \,\,\, F_{h_1}^{(l_1)}\dots F_{h_k}^{(l_k)}  
\eeq
means in fact a sum of $(2l-1)!!$ terms containing all the possible pairings of $2l$ indices of the matrix $\kappa$,  with the $2l$ indices of the tensors $F_{h_i}^{(l_i)}$.

\medskip

For example to order $N^{-2}$, i.e. $g=2$ we have:
\beq
\hat{F}_2 = F_2 + \kappa\left({F_1''\over 2} + {(F_1')^2\over 2} \right)
+ \,3\kappa^2\left({F_0''''\over 4!}  + 2\, { F_1'\, F_0'''\over 2\,\,3!}  \right)
+ 15\, \kappa^3 \left( {(F_0''')^2\over 2\,\, 3! \, 3!}   \right)
\eeq
where the last term $ 15\, \kappa^3 \, (F_0''')^2$ contains two topologically inequivalent types of pairings among the indices:
\bea
15\, \kappa^3 \,\, (F_0''')^2
&\to &
\sum_{i_1,i_2,i_3,i_4,i_5,i_6}
9\,\,\, \kappa_{i_1,i_2} \kappa_{i_3,i_4}\kappa_{i_5,i_6}\,\,
{\partial^3 F_0\over \partial \epsilon_{i_1}\partial \epsilon_{i_2}\partial \epsilon_{i_3}}\,\,
{\partial^3 F_0\over \partial \epsilon_{i_4}\partial \epsilon_{i_5}\partial \epsilon_{i_6}} \cr
&& +
6\,\,\, \kappa_{i_1,i_4} \kappa_{i_2,i_5}\kappa_{i_3,i_6}\,\,
{\partial^3 F_0\over \partial \epsilon_{i_1}\partial \epsilon_{i_2}\partial \epsilon_{i_3}}\,\,
{\partial^3 F_0\over \partial \epsilon_{i_4}\partial \epsilon_{i_5}\partial \epsilon_{i_6}}
\eea
This equation can be diagrammatically represented as follows \cite{abk}:
\bea
\hat{F}_2
&=&
{\mbox{\epsfxsize=1.5truecm\epsfbox{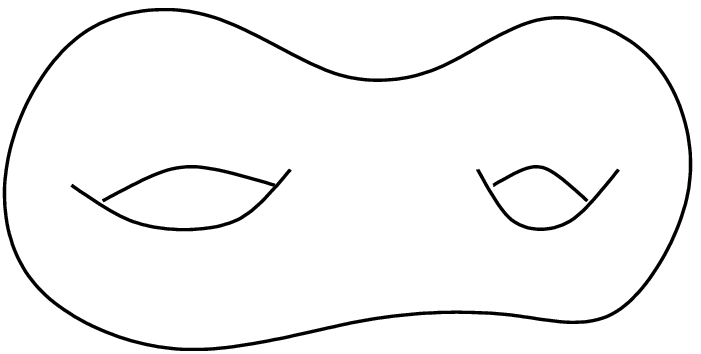}}}
+{1\over 2}{\mbox{\epsfxsize=1truecm\epsfbox{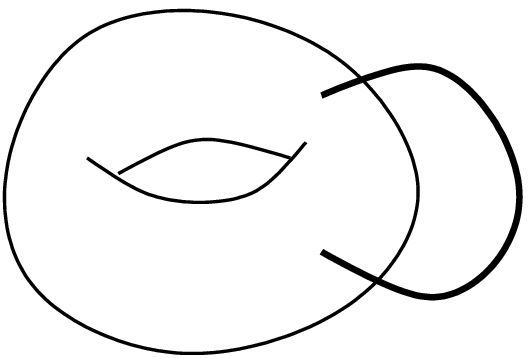}}}
+{1\over 2}{\mbox{\epsfxsize=2truecm\epsfbox{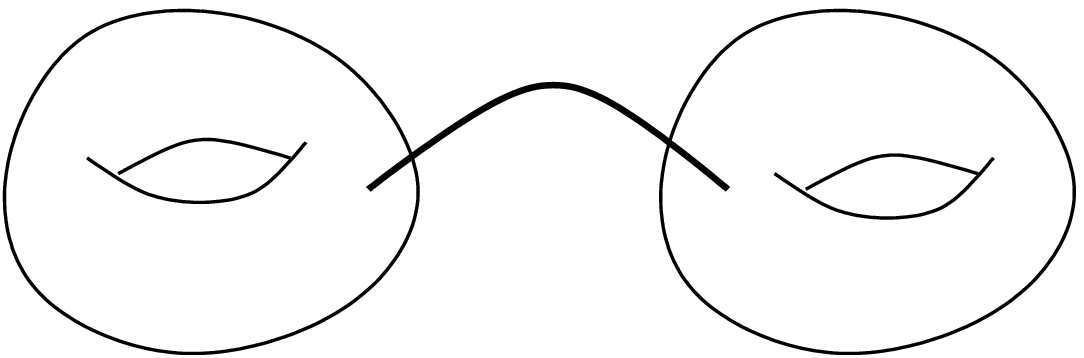}}}
+{1\over 8}{\mbox{\epsfxsize=1.2truecm\epsfbox{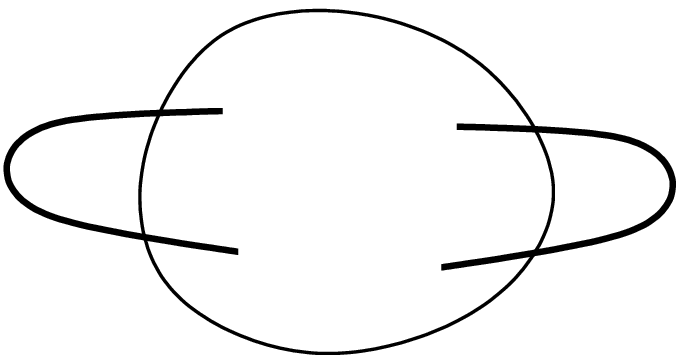}}}
+{1\over 2}{\mbox{\epsfxsize=2.truecm\epsfbox{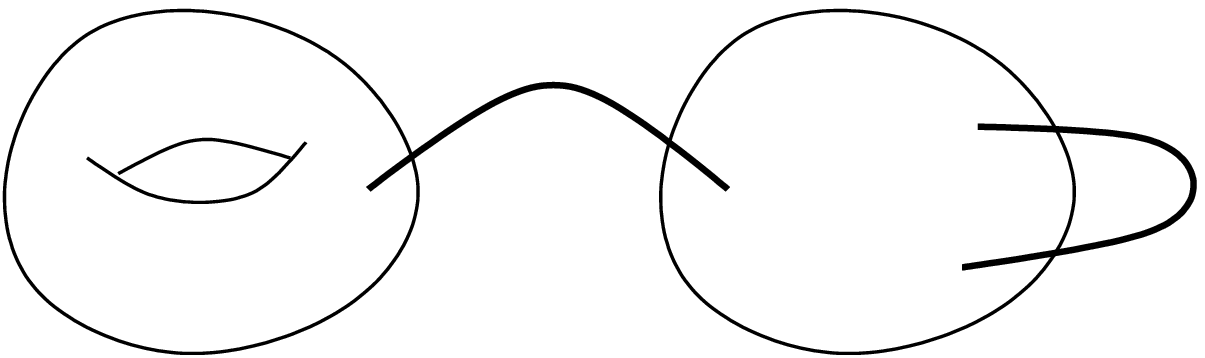}}} \cr
&& +{1\over 8}{\mbox{\epsfxsize=2.2truecm\epsfbox{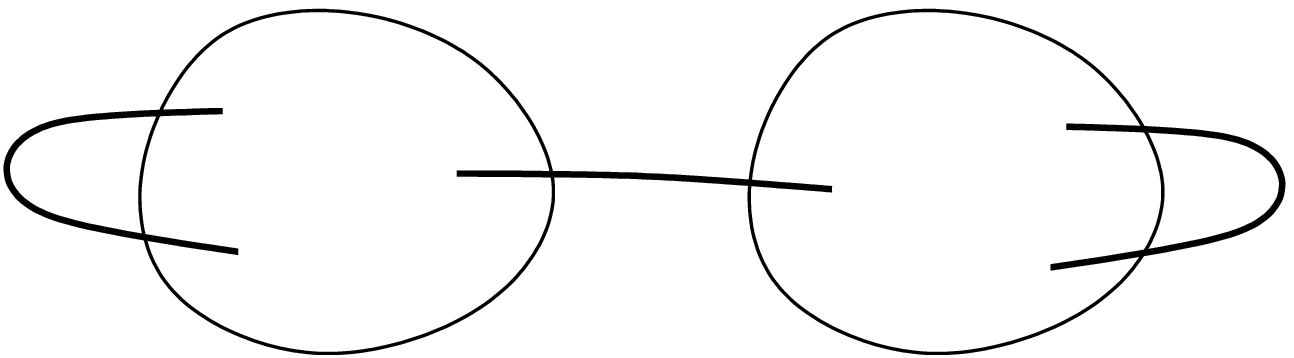}}}
+{1\over 12}{\mbox{\epsfxsize=2.2truecm\epsfbox{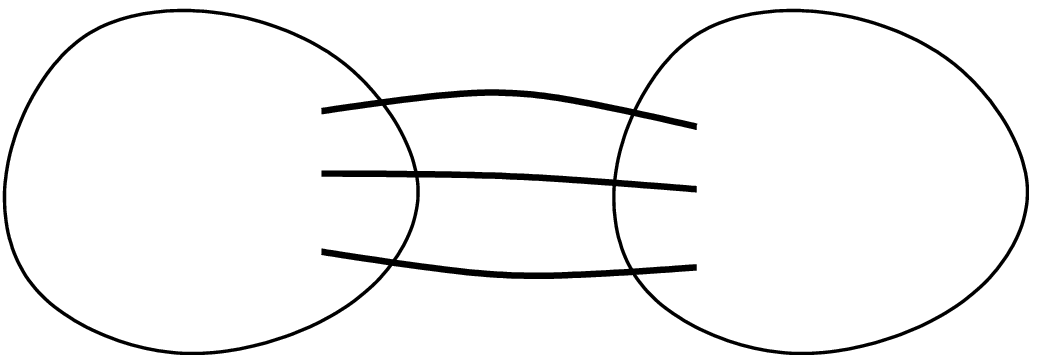}}}
\eea
where each term represents a possible degeneracy of a genus 2 Riemann surface (imagine each link   contracted to a point).
The prefactor is  $1/\# Aut$, i.e. the inverse of the number of automorphisms, for instance in the last graph we have a ${\mathbb Z}_2$ symmetry by exchanging the 2 spheres, and a $\sigma_3$ symmetry from permuting the 3 endpoints of the edges, i.e. $12=\#({\mathbb Z}_2\times \sigma_3)$ automorphisms.

\bigskip

Formally, eq.\ref{eqZhatkappa} is very similar to eq.\ref{exphatZTheta}, with the identification:
\beq\label{kappaTheta}
(2k-1)!!\,\, \kappa^{k} \to \, \Theta^{(2k)}
\eeq

{\bf proof:}
eq.\ref{eqZhatkappa} is the Wick theorem expansion of the following integral \cite{abk, eynHolan}:
\bea
Z(\bfeps^*,\kappa) 
&=&  \ee{\sum_h N^{2-2h} F_h(\bfeps^*,\kappa)}  \cr
&=& \int d\eta\, \ee{F(\eta) - N^2 (\eta-\bfeps^*) F_0' - {N^2\over 2} (\eta-\bfeps^*) F_0''(\eta-\bfeps^*) - {N^2 i \pi}(\eta-\bfeps^*) \kappa^{-1} (\eta-\bfeps^*) }  \cr
&=& Z(\bfeps^*)\, \int d\eta\, \ee{ \sum_{l>0} \sum_{h>1-l/2} {N^{2-2h}\over l!}\, F_h^{(l)} (\eta-\bfeps^*)^l  - {N^2  i \pi }(\eta-\bfeps^*) \kappa^{-1} (\eta-\bfeps^*) }  \cr
&=& Z(\bfeps^*)\, \sum_k \sum_{l_i>0} \sum_{h_i>1-l_i/2} {N^{\sum_i 2-2h_i}\over k! l_1!\dots l_k!}
F_{h_1}^{(l_1)}\dots F_{h_k}^{(l_k)}\,\cr
&& \qquad  \int d\eta \,\, (\eta-\bfeps^*)^{\sum l_i}  \,\,\ee{- {N^2 i \pi}(\eta-\bfeps^*) \kappa^{-1} (\eta-\bfeps^*) }  
\eea
i.e.
\bea
Z(\bfeps*,\kappa)
&=&  Z(\bfeps^*)\,  \sum_{k} \sum_{l_i}\sum_{h_i} {N^{\sum_i (2-2h_i-l_i)}\over k! l_1!\,\dots\, l_k!} F_{h_1}^{(l_1)}\dots F_{h_k}^{(l_k)}  \,\, \left.{\partial^{\sum l_i} f\over \partial u^{\sum l_i}}\right|_{u=0,t=-{1\over 2}\kappa^{-1}}
\cr
\eea
where $f(u,t)$ is nearly the same as $\Theta$ except that we have an integral instead of a sum over integers:
\bea
f(u,t) &=& \int d\bfeps\,\, \ee{N(\bfeps-\bfeps^*)u}\,\,\ee{N^2 (\bfeps-\bfeps^*)t(\bfeps-\bfeps^*)}\,\,\ee{2i\pi N\, \bfeps\nu} \cr
&=& \ee{2i\pi N\, \bfeps^*\nu}\,\int d\bfeps\,\, \ee{N \bfeps (u+2i\pi\nu)}\,\,\ee{N^2 \bfeps t \bfeps} \cr
&=& \ee{2i\pi N\, \bfeps^*\nu}\,\ee{-{1\over 4}\,(u+2i\pi\nu) t^{-1} (u+2i\pi\nu)} \cr
\eea
It also satisfies:
\beq
{\partial f\over \partial t} = {\partial^2 f\over \partial u^2}
\eeq
It is clear that:
\beq
\left.{\partial^{2l+1} f\over \partial u^{\sum l_i}}\right|_{u=0,t=-{1\over 2}\kappa^{-1}}
= 0
\virg
\left.{\partial^{2l} f\over \partial u^{\sum l_i}}\right|_{u=0,t=-{1\over 2}\kappa^{-1}}
= (2l-1)!!\,\, \kappa^l
\eeq
which proves our claim eq.\ref{kappaTheta}.

\medskip

This analogy between convergent integrals obtained by summing over filling fractions, and holomorphic anomaly equations is puzzling, and it would be worth getting some understanding of that fact.

\section{Background independence}

So far, $\bfeps^*$ was chosen arbitrary, and eq.\ref{exphatZTheta}, eq.\ref{ZThetaresummed} and the property \ref{kappaTheta} hold independently of the choice of $\bfeps^*$. 
Indeed $\hat{Z}(\gamma)$ does not depend at all on a choice of $\bfeps^*$.

If we take eq.\ref{exphatZTheta} as a definition of a string theory partition function, it seems at first sight that it depends on $\bfeps^*$, but in fact it does not.
Those facts are related to the so-called "background independence" problem in string theory \cite{Witten}.

\bigskip

From \cite{Bertasympt}, it should be expected that if we choose $\bfeps^*$ such that the spectral curve has the Boutroux property:
\beq
{\rm Boutroux\,\, property:}\qquad \quad \forall {\cal C}\, , \qquad \Re\,\oint_{{\cal C}} y dx =0
\eeq
then, the formal series $\sum_g N^{2-2g} F_g$ as well as the $\Theta$-sums in eq.\ref{exphatZTheta} and  eq.\ref{ZThetaresummed}, should be convergent series, and thus we really have an asymptotic expansion instead of only an asymptotic series.
However, this fact is not proved yet (except for the 1-matrix model).

Boutroux curves in particular, are such that:
\beq\label{Boutroux}
\bfeps^*={1\over 2i\pi}\oint_{\acycle} ydx \in {\mathbb R}^\genus
\virg
\Re\, F'_0 = \Re\oint_{\bcycle} ydx = 0
\eeq
Boutroux curves can be obtained as follows:
Notice that $\Re F_0'' = -\pi\, \Im\tau<0$ (the imaginary part of the Riemann matrix of periods is always positive, see \cite{Farkas, Fay}), and thus $-\Re F_0$ is a convex function on $\bfeps^* \in {\mathbb R}^\genus$, therefore it has a minimum in each cell of the moduli space. The minimum clearly satisfies eq.\ref{Boutroux}.
In other words there should be one Boutroux curve in each cell of the moduli space of the spectral curve.
One may expect that each cell corresponds to a possible connectivity pattern of the generalized path $\gamma$.

Notice that if the potentials are real, and the filling fraction $\bfeps^*$ is real, then $F_0$ is real as well, and the Boutroux condition becomes $F'_0=0$.

\section{Conclusion}

In this article, we have improved the asymptotic (conjectured) formula of \cite{BDE} for matrix integrals to all orders.
We have also found an interesting connection between this expansion and combinatoric geometry of degenerate Riemann surfaces, through the holomorphic anomaly equation.

\medskip
The relationship between higher genus $\genus>0$ formal matrix integrals and nodal discrete surfaces  was already known \cite{eynform, BDE}, and here we see that there is also a relationship with nodal Riemann surfaces.
In fact, so far all intersection numbers in Kontsevich integral \cite{EOFg}, or Weil-Petersson volumes \cite{eynOWP,eynkappa}, were computed with genus zero ($\genus=0$) spectral curve formal matrix models. This works tends to show that higher genus spectral curves have to do with nodal surfaces.
This relationship needs to be further investigated.

\section*{Acknowledgments}
We would like to thank M. Mari\~no for careful reading of the manuscript, and M. Bertola, T. Grava, N. Orantin for useful and fruitful discussions on this subject.
This work is partly supported by the Enigma European network MRT-CT-2004-5652, by the ANR project G\'eom\'etrie et int\'egrabilit\'e en physique math\'ematique ANR-05-BLAN-0029-01, by the Enrage European network MRTN-CT-2004-005616,
by the European Science Foundation through the Misgam program,
by the French and Japaneese governments through PAI Sakurav, by the Quebec government with the FQRNT.

\end{document}